\documentclass{mem}
\usepackage{natbib}
\usepackage{txfonts}
\usepackage{graphicx}
\usepackage{psfig}
\usepackage[a4paper,breaklinks,dvipdfm]{hyperref}
\idline{75}{282}
\begin{document}
\def\teff{$T\rm_{eff }$}
\def\kms{$\mathrm {km s}^{-1}$}

\title{
Overionization in X-ray spectra: a new paradigm for Mixed-Morphology SNRs
}

   \subtitle{}

\author{
M. \,Miceli\inst{1,2} 
          }

  \offprints{M. Miceli}

\institute{
Dipartimento di Fisica, Universit\`a di Palermo, Piazza del Parlamento 1, 90134 Palermo, Italy
\and
INAF-Osservatorio Astronomico di Palermo, Piazza
del Parlamento 1, 90134 Palermo, Italy
\email{miceli@astropa.unipa.it}
}

\authorrunning{Miceli}

\titlerunning{Overionization in MM SNRs}

\abstract{
Mixed-morphology SNRs are characterized by a shell-like radio emission, a centrally peaked X-ray morphology, and by interaction with molecular clouds. Many models have been proposed to explain these peculiar remnants, but their physical origin is still unclear. The recent discovery of overionized (i. e. recombining) ejecta in 3 mixed-morphology SNRs has dramatically challenged all the previous models and opened up new, unexpected scenarios. I review the main properties of these remnants and their peculiar X-ray spectral properties. I also discuss the hydrodynamic model developed to explain the presence of overionized ejecta in W49B and present a list of open issues that still need to be clarified.

}
\maketitle{}

\section{Introduction}

Supernova Remnants (SNRs) have been observed in X-rays for decades and spatially resolved spectral analysis allowed us to ascertain a lot of important information about the physics of the post-shock high-temperature plasma and the mechanisms of particle acceleration.

X-ray spectra from SNRs are typically composed by i) a non-thermal component, associated with synchrotron emission from TeV electrons accelerated at the shock front and dominating the spectra of young remnants (with shock velocity of a few $10^3$ km$/$s) and ii) a thermal component associated with the optically thin interstellar medium (ISM) heated by the main shock front and$/$or with the ejecta heated by the reverse shock front. 
The prompt heating of the plasma at the shock front is followed by a slower process of ionization (through ion-electron collisions), where ions tend to reach the charge state distribution that is due to their post-shock temperature. The Collisional Ionization Equilibrium (CIE) is reached when the ionization parameter $\tau_{NEI}=\int ndt\ga10^{12}$ s cm$^{-3}$ (see, for example, \citealt{vb99}). Assuming a typical density of 1 cm$^{-3}$, we derive that the post-shock plasma in remnants younger than $\sim3\times10^4$ yr is expected to be underionized. Underionization is indeed a typical feature of X-ray spectra and has been observed in many cases, especially in young SNRs, like Cas A (\citealt{hlb04}, \citealt{lds06}), Kepler (\citealt{cdb04}), SN~1006 (\citealt{abd07}, \citealt{mbi09}).

The ``canonical" scenario of underionized plasma in SNRs has been severely challenged by the recent discovery of overionized (i. e. recombining) plasma in three SNRs, namely IC~443, W49B, and G359.1$-$0.5 (see Sect. \ref{over}). All these remnants belong to the class of Mixed-Morphology SNRs (MMSNRs) and are interacting with nearby molecular clouds. 

In this paper I briefly review the main properties of MMSNRs, their relation with molecular clouds, and the models proposed to explain their puzzling morphology (Sect. \ref{MM}). I then focus on the presence of overionized ejecta (Sect. \ref{over}) and discuss its possible physical origin. The conclusions and the open issues are listed in Sect. \ref{Conclusions}

\section{Mixed-morphology supernova remnants}
\label{MM}
Core-collapse supernovae are expected to explode in the same dense and complex environment where their progenitors were born. In fact, massive stars cannot drift far away from the star-forming region because of their very short life-times. Moreover, massive stars strongly modify the circumstellar medium (CSM) during their post main-sequence evolution, through strong stellar winds that can create tenuous wind-blown bubbles enclosed by dense cavity walls. SNRs can interact with wind residuals and impact, during their evolution, on the cavity wall, as modelled by \citet{dwa05,dwa07}.
We therefore expect to find several SNRs interacting with molecular clouds (MCs). These interactions are very important since they drive the exchange of energy and mass in the galactic ISM.
There are several examples of galactic SNRs interacting with a cavity wall and/or molecular clouds (a list of these remnants is provided by \citealt{jcw10}\footnote{See also \newline\scriptsize{http://astronomy.nju.edu.cn/~ygchen/others/bjiang/interSNR6.htm})}). 

\begin{figure}[htb!]
 \centerline{\hbox{     
     \psfig{figure=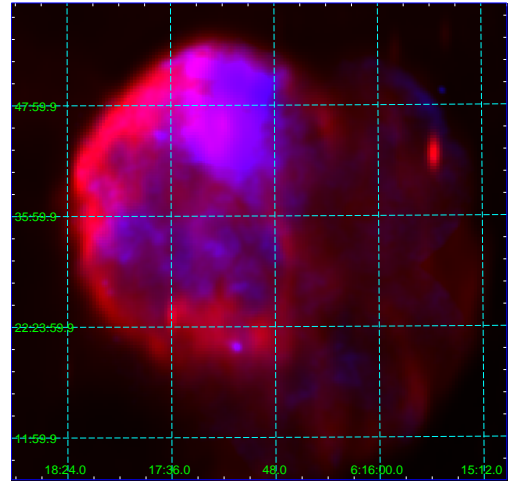,width=\columnwidth}
     }}
\caption{Radio (1420 MHz, in red) and X-ray ($0.5-2$ keV, in blue) composite image of IC 443, an example of mixed-morphology SNR. Details on the analysis of the radio and X-ray data are reported in \citet{lea04b} and \citet{tbr06}, respectively.}
\label{fig:ic443}
\end{figure}

MMSNRs (\citealt{rp98}, hereafter RP98) are typically interacting with molecular or H I clouds. They are characterized by a shell-like morphology in the radio band and a centrally filled emission in X-rays, with little or no limb brightening. Figure \ref{fig:ic443} shows, as an example of MMSNR, the X-ray and radio images of IC 443. At odds with plerionic SNRs, whose X-ray emission is associated with the central pulsar wind nebula, MMSNRs have a thermal component in their spectra and the centrally peaked X-ray emission cannot be explained in the framework of the Sedov model (\citealt{sed59}).

Different models have been proposed to explain the puzzling morphology of these sources. In particular, \citet{wl91} modelled MMSNRs as remnants expanding in a cloudy environment, and associated the centrally bright X-ray emission with small ISM cloudlets engulfed by the main shock front and heated up to X-ray emitting temperatures by thermal conduction with the hot, tenuous interior of the SNR (i.e. the clouds ``evaporate" in the hotter intercloud medium). In this model the density and temperature radial profiles depend on the parameter $C/\tau$, where $C$ is the ratio of the mass in clouds to the mass in the intercloud medium and $\tau$ is the ratio of the characteristic evaporation time-scale of the clouds to the age of the remnant.

A second model has been proposed by \citet{csm99}, who considered a remnant evolving in an ambient medium of fairly high density ($\la5$ cm$^{-3}$) and included the effects of thermal conduction. Because of the high density, after a few 10$^4$ yr, radiative cooling plays an important role in the post-shock plasma, whose temperature drops well below $10^6$ K (therefore the shell does not emit in X-rays). The deep interior of the remnant has a substantial and fairly uniform pressure, and thermal conduction prevents formation of the rarefied cavity characteristic of Sedov evolution. This effect increases the emission measure in the center of the remnant with respect to that predicted by the Sedov solution and could explain the centrally bright X-ray emission of MMSNRs.

However, \citet{bmt09} showed that these models fail in reproducing the observed features of the MMSNRs IC 443 and G166.0+4.3. Also, \citet{ls06} showed that the Cox model requires very high gas density that are not consistent with those observed. Both the Cox and the White \& Long models associate the bulk of the X-ray emission of MMSNRs with the shocked ISM (as in the original classification by RP98), while, as first proposed by \citet{ssh02}, ejecta emission can be non-negligible. Indeed, it has been shown that in many MMSNRs ($\sim50\%$) X-ray emission from the ejecta is significant (\citealt{ls06}, \citealt{tbm08}, \citealt{bmt09}), and in some cases it dominates (e.~g. W49B). Therefore, it is necessary to include an accurate model of the ejecta evolution to understand the peculiar morphology of these remnants.

\section{Overionized ejecta in MMSNRs}
\label{over}

A new, unexpected, discovery has recently involved the X-ray emission from ejecta in MMSNRs: spectral analysis carried on with the $Suzaku$ satellite has shown the presence of overionized material. 

The first detection of recombining Si-rich and S-rich plasma has been obtained by \citet{yok09} in the MMSNR IC 443. The spectrum of this remnant shows characteristic edge-like features associated with free-bound transitions of electrons to the K-shell of H-like Si and S ions. The flux of Radiative Recombination Continuum (RRC) is a significant fraction of the flux in the He-like K$\alpha$ line ($\sim 0.28$ and $\sim0.18$ for Si and S, respectively) and is much higher than that predicted in CIE, thus proving that the level of ionization is higher than that expected at the observed (electron) temperature.

Overionized ejecta were observed also in W49B by \citet{oky09}, who detected a hard RRC emerging above 8 keV and associated with recombination of H-like Fe ions into the ground state of He-like ions (dominant contribution) and of fully ionized Fe ions into the ground state of H-like ions.

The third example of MMSNR with overionized ejecta is G359.1-0.5 (\citealt{okt11}). In this case a spectral model with a singular component of overionized plasma with enhanced metallicity (ejecta) can fit the global spectrum, thus providing information on the electron temperature, $kT_e\sim0.29$ keV, and on the ionization temperature (i.~e. the temperature corresponding to the observed charge state distribution in CIE), $kT_z\sim0.77$ keV. This result clearly shows that ejecta have been heated up to very high temperatures and are now rapidly cooling down.

Prompted by these results, new $Suzaku$ observations of MMSNRs have been performed and the preliminary results are encouraging, showing new cases of overionized ejecta (\citealt{koy08}). The presence of recombining ejecta may, therefore, be a common feature of MMSNRs, though the relationship between the centrally peaked X-ray morphology and overionization is really puzzling.

\begin{figure}[htb!]
 \centerline{\hbox{     
     \psfig{figure=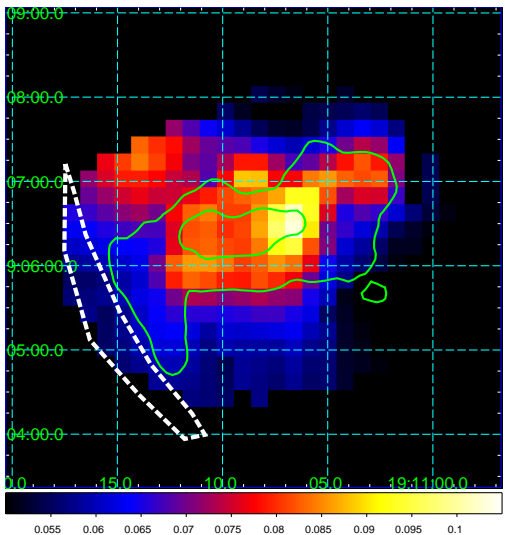,width=\columnwidth}
     }}
\caption{Vignetting-corrected and background-subtracted \emph{XMM-Newton} pn map of the ratio of the count-rate in the RRC ($8.3-12$ keV) and bremsstrahlung ($4.4-6.2$ keV) bands in W49B. Higher ratios indicate stronger overionization (for details see \citealt{mbd10}). The white dashed region indicate the approximate position of the molecular cloud interacting with the remnant, and the green contour lines mark the \emph{XMM-Newton} EPIC count-rate (at $33\%$ and $66\%$ of the maximum) in the $1-9$ keV band.}
\label{fig:w49b}
\end{figure}

In order to ascertain the physical origin of this rapidly cooling plasma, it is important to focus on its spatial localization within the X-ray emitting ejecta. A first attempt in this direction has been performed for W49B (\citealt{mbd10}). This remnant shows a jet-like morphology in the X-ray band, with jet axis oriented along the East-West direction and the eastern jet being confined by a dense molecular cloud. In W49B the overionized plasma is localized in the center of the remnant and in its western jet, while it is not detected in the bright eastern jet, where the expansion of the ejecta is hampered by their interaction with the dense cloud. Figure \ref{fig:w49b} shows the ratio of the RRC to the bremsstrahlung count-rates (that is a proxy of the overionized plasma). Since overionization is present only where ejecta can expand freely, it is possible that the rapid cooling that generates the overionization is due to the adiabatic ejecta expansion. 
Nevertheless, an early heating is also necessary and a possible mechanism for this process can be the interaction of the remnant with nearby circumstellar material (e.~g. wind residuals from the progenitor star).

This scenario has been quantitatively investigated by \citet{zmb11} who performed detailed hydrodynamic simulations (taking into account radiative losses and thermal conduction) of W49B, including in their modeling an accurate description of the circumstellar environment and deviations from ionization equilibrium induced by plasma dynamics. They found that the early interaction of the remnant with dense circumstellar rings (observed by \citealt{krr07}) can heat the Fe ions up to very high temperatures ($>10$ keV) and that subsequent  adiabatic expansion can reproduce the observed distribution of overionized ejecta. This model also naturally reproduces the peculiar jet-like morphology of the X-ray emission.

\section{Conclusions}
\label{Conclusions}

Recent X-ray observations have discovered overionized plasma in IC443, W49B, and G359.1-0.5 and indicate that overionization may be a common feature of MMSNRs with X-ray emitting ejecta.
An X-ray emitting overionized plasma is the result of an early heating followed by a rapid cooling, but the mechanisms responsible for such processes in MMSNRs are still not completely understood.

The localization of the overionized plasma in W49B has provided important information and detailed hydrodynamic modeling has shown that the interaction of the remnant with nearby CSM can provide an early heating of the ejecta followed by their rapid adiabatic expansion. This mechanism can naturally explain the observed features of W49B, but up to now it is not possible to extend this picture to other MMSNRs.

Nevertheless, the association of MMSNRs with molecular clouds supports a core-collapse origin for these remnants and this makes the presence of nearby, dense CSM likely. As in W49B, the reflected shocks originating in the interaction of the forward shock with the CSM may provide an efficient mechanism to heat the ejecta in the early phases of the remnant evolution. 

To verify this hypothesis a detailed (spatially resolved) spectral analysis of MMSNRs is necessary. Deep observations are required and spectral models with complete description of RRC and cascade effects are necessary to obtain reliable diagnostics. Also hydrodynamic models that include realistic description of the environment around these remnants (together with NEI effects) are necessary.

In conclusion, the discovery of overionized plasma has opened up new scenarios to explain the puzzling nature of MMSNRs and is pushing forward the development of new, accurate spectral codes and new hydrodynamic models necessary to correctly interpret the observation results.

\begin{acknowledgements}
I wish to thank the organizers, especially Alexandre Marcowith, for their kind invitation to this very interesting and fruitful conference. 
\end{acknowledgements}

\bibliographystyle{aa}

\begin{thebibliography}{26}
\expandafter\ifx\csname natexlab\endcsname\relax\def\natexlab#1{#1}\fi

\bibitem[{{Acero} {et~al.}(2007){Acero}, {Ballet}, \& {Decourchelle}}]{abd07}
{Acero}, F., {Ballet}, J., \& {Decourchelle}, A. 2007, \aap, 475, 883

\bibitem[{{Bocchino} {et~al.}(2009){Bocchino}, {Miceli}, \& {Troja}}]{bmt09}
{Bocchino}, F., {Miceli}, M., \& {Troja}, E. 2009, \aap, 498, 139

\bibitem[{{Cassam-Chena{\"i}} {et~al.}(2004){Cassam-Chena{\"i}},
  {Decourchelle}, {Ballet}, {Hwang}, {Hughes}, {Petre}, \& {et al.}}]{cdb04}
{Cassam-Chena{\"i}}, G., {Decourchelle}, A., {Ballet}, J., {et~al.} 2004, \aap,
  414, 545

\bibitem[{{Cox} {et~al.}(1999){Cox}, {Shelton}, {Maciejewski}, {Smith},
  {Plewa}, {Pawl}, \& {R{\' o}{\. z}yczka}}]{csm99}
{Cox}, D.~P., {Shelton}, R.~L., {Maciejewski}, W., {et~al.} 1999, \apj, 524,
  179

\bibitem[{{Dwarkadas}(2005)}]{dwa05}
{Dwarkadas}, V.~V. 2005, \apj, 630, 892

\bibitem[{{Dwarkadas}(2007)}]{dwa07}
{Dwarkadas}, V.~V. 2007, \apj, 667, 226

\bibitem[{{Hwang} {et~al.}(2004){Hwang}, {Laming}, {Badenes}, {Berendse},
  {Blondin}, {Cioffi}, {DeLaney}, {Dewey}, {Fesen}, {Flanagan}, {Fryer},
  {Ghavamian}, {Hughes}, {Morse}, {Plucinsky}, {Petre}, {Pohl}, {Rudnick},
  {Sankrit}, {Slane}, {Smith}, {Vink}, \& {Warren}}]{hlb04}
{Hwang}, U., {Laming}, J.~M., {Badenes}, C., {et~al.} 2004, \apjl, 615, L117

\bibitem[{{Jiang} {et~al.}(2010){Jiang}, {Chen}, {Wang}, {Su}, {Zhou},
  {Safi-Harb}, \& {DeLaney}}]{jcw10}
{Jiang}, B., {Chen}, Y., {Wang}, J., {et~al.} 2010, \apj, 712, 1147

\bibitem[{{Keohane} {et~al.}(2007){Keohane}, {Reach}, {Rho}, \&
  {Jarrett}}]{krr07}
{Keohane}, J.~W., {Reach}, W.~T., {Rho}, J., \& {Jarrett}, T.~H. 2007, \apj,
  654, 938

\bibitem[{{Koyama}(2010)}]{koy08}
{Koyama}, K. 2010, in 38th COSPAR Scientific Assembly, Vol.~38, 2773--+

\bibitem[{{Lazendic} {et~al.}(2006){Lazendic}, {Dewey}, {Schulz}, \&
  {Canizares}}]{lds06}
{Lazendic}, J.~S., {Dewey}, D., {Schulz}, N.~S., \& {Canizares}, C.~R. 2006,
  \apj, 651, 250

\bibitem[{{Lazendic} \& {Slane}(2006)}]{ls06}
{Lazendic}, J.~S. \& {Slane}, P.~O. 2006, \apj, 647, 350

\bibitem[{{Leahy}(2004)}]{lea04b}
{Leahy}, D.~A. 2004, \aj, 127, 2277

\bibitem[{{Miceli} {et~al.}(2010){Miceli}, {Bocchino}, {Decourchelle},
  {Ballet}, \& {Reale}}]{mbd10}
{Miceli}, M., {Bocchino}, F., {Decourchelle}, A., {Ballet}, J., \& {Reale}, F.
  2010, \aap, 514, L2+

\bibitem[{{Miceli} {et~al.}(2009){Miceli}, {Bocchino}, {Iakubovskyi},
  {Orlando}, {Telezhinsky}, {Kirsch}, {Petruk}, {Dubner}, \&
  {Castelletti}}]{mbi09}
{Miceli}, M., {Bocchino}, F., {Iakubovskyi}, D., {et~al.} 2009, \aap, 501, 239

\bibitem[{{Ohnishi} {et~al.}(2011){Ohnishi}, {Koyama}, {Tsuru}, {Masai},
  {Yamaguchi}, \& {Ozawa}}]{okt11}
{Ohnishi}, T., {Koyama}, K., {Tsuru}, T.~G., {et~al.} 2011, \pasj, 63, 527

\bibitem[{{Ozawa} {et~al.}(2009){Ozawa}, {Koyama}, {Yamaguchi}, {Masai}, \&
  {Tamagawa}}]{oky09}
{Ozawa}, M., {Koyama}, K., {Yamaguchi}, H., {Masai}, K., \& {Tamagawa}, T.
  2009, \apjl, 706, L71

\bibitem[{{Rho} \& {Petre}(1998)}]{rp98}
{Rho}, J. \& {Petre}, R. 1998, \apjl, 503, L167

\bibitem[{{Sedov}(1959)}]{sed59}
{Sedov}, L.~I. 1959, Similarity and Dimensional Methods in Mechanics (New York:
  Academic Press, 1959)

\bibitem[{{Slane} {et~al.}(2002){Slane}, {Smith}, {Hughes}, \& {Petre}}]{ssh02}
{Slane}, P., {Smith}, R.~K., {Hughes}, J.~P., \& {Petre}, R. 2002, \apj, 564,
  284

\bibitem[{{Troja} {et~al.}(2008){Troja}, {Bocchino}, {Miceli}, \&
  {Reale}}]{tbm08}
{Troja}, E., {Bocchino}, F., {Miceli}, M., \& {Reale}, F. 2008, \aap, 485, 777

\bibitem[{{Troja} {et~al.}(2006){Troja}, {Bocchino}, \& {Reale}}]{tbr06}
{Troja}, E., {Bocchino}, F., \& {Reale}, F. 2006, \apj, 649, 258

\bibitem[{{van Paradijs} \& {Bleeker}(1999)}]{vb99}
{van Paradijs}, J. \& {Bleeker}, J.~A.~M., eds. 1999, Lecture Notes in Physics,
  Berlin Springer Verlag, Vol. 520, {X-Ray Spectroscopy in Astrophysics}

\bibitem[{{White} \& {Long}(1991)}]{wl91}
{White}, R.~L. \& {Long}, K.~S. 1991, \apj, 373, 543

\bibitem[{{Yamaguchi} {et~al.}(2009){Yamaguchi}, {Ozawa}, {Koyama}, {Masai},
  {Hiraga}, {Ozaki}, \& {Yonetoku}}]{yok09}
{Yamaguchi}, H., {Ozawa}, M., {Koyama}, K., {et~al.} 2009, \apjl, 705, L6

\bibitem[{{Zhou} {et~al.}(2011){Zhou}, {Miceli}, {Bocchino}, {Orlando}, \&
  {Chen}}]{zmb11}
{Zhou}, X., {Miceli}, M., {Bocchino}, F., {Orlando}, S., \& {Chen}, Y. 2011,
  \mnras, 415, 244

\end{thebibliography}

\end{document}